\documentclass[aps,showpacs]{revtex4}
\usepackage{amssymb}
\usepackage{graphics}
\usepackage{epsfig}

\begin{document}

\newcommand{\nb}{\nonumber}

\title{ Prompt heavy quarkonium production in association with a massive
(anti)bottom quark at the LHC}

\author{ Li Gang, Wang ShuangTe, Song Mao, Lin JiPing   \\
{\small School of Physics and Material Science, Anhui University,
Hefei, Anhui 230039, P.R.China} }

\begin{abstract}
In this work, we investigate the associated production of prompt
heavy quarkonium and a massive (anti)bottom quark to leading order
in the nonrelativistic factorization formalism at the LHC. We present
numerical results for the processes involving
$J/\psi,\chi_{cJ},\Upsilon$ and $\chi_{bJ}$. From our work, we find
that the production rates of these processes are quite large, and
these processes have the potential to be detected at the LHC. When
$p_T$ is smaller than about 10 GeV, the $c\bar{c} [ ^1S_0^{(8)} ]$
state give the main contribution to the $p_T$ distribution of prompt
$J/\psi$ with a massive (anti)bottom quark production. For the process
$pp \to \Upsilon+b(\bar{b})$, the contribution of the color-singlet mechanism is larger
than that in the color-octet mechanism at low $p_T$ region. We also investigate the
processes $pp\to \chi_{cJ}+b(\bar{b})$ and $pp \to
\chi_{bJ}+b(\bar{b})$, the $p_T$ distributions
are dominated by the color-octet Fock state contribution at large $p_T$.
These processes provide an interesting signature that could
be studied at the LHC, and the measurement of these processes is
useful to test the color-singlet mechanism and color-octet mechanism.
\end{abstract}

\pacs{12.38.Bx, 13.60.Le, 13.85.Ni, 14.40.Pq} \maketitle

\newpage
\section{Introduction}
A heavy quarkonium is made of a heavy quark and a heavy antiquark;
its decay and production can be factorized into a short-distance
part, which can be calculated in QCD perturbatively, and a
long-distance part, which is governed by nonperturbative QCD
dynamics. So the study of heavy quarkonium offers a good
testing ground for investigating of strong interaction in both
perturbative and nonperturbative level in high-energy physics.

In the early days, based on the factorization formalism,
color-singlet mechanism(CSM) \cite{csm} was used to describe the
production and decay of heavy quarkonium, and obtained some
phenomenological successes. However, the CSM has encountered many
difficulties in various theoretical \cite{Barbieri:1976fp} and
experimental aspects \cite{Abe:1992ww}, such as the appearance of a
logarithmic infrared divergence in the case of (next-to-leading-order)NLO $P$-wave decays
into light hadrons and the huge discrepancy of the high-$p_T$
$J/\psi$ production between the theoretical prediction and the
experimental measurement at the Tevatron. To solve these formal and
phenomenological problems of CSM, a rigorous theory, nonrelativistic
QCD(NRQCD) \cite{bbl} was proposed by Bodwin, Braaten and
Lepage(BBL). In the NRQCD, the idea of perturbative factorization is
retained, the processes of production and decay of heavy quarkonium are
also separated into two parts: a short-distance part and long-distance
matrix elements (LDMEs), which can be extracted from experiments.
The relative importance of the LDMEs can be estimated by means of
the velocity-scaling rules \cite{vrule}.

The vital difference between NRQCD and the traditional color-singlet model is that,
in NRQCD, the complete structure of the quarkonium Fock space has been explicitly
considered, and NRQCD predicts the existence of color-octet (CO)
processes in nature. This CO theory allows that $Q\bar{Q}$ pairs can
be produced at short distance in CO states and subsequently evolved
into physical, color-singlet (CS) quarkonium through emission of
soft gluons at nonperturbative process \cite{co2cs}. Being
introduced to the color-octet mechanism (COM), NRQCD has absorbed
the infrared divergences in P-wave \cite{bbl,p1,p2} and
D-wave \cite{d1,d2} decay widths of heavy quarkonium and
successfully reconciled the orders of magnitude discrepancy
between the experimental data of $J/\psi$ production at the Tevatron
\cite{comtev} and the leading order (LO) CSM theoretical predictions
at large $p_T$.

 \par
Recently, in order to clarify the validity and limitation of the NRQCD formalism,
substantial progress has been achieved in the calculations of heavy quarkonium production.
The DELPHI data favor the NRQCD COM prediction for the $\gamma\gamma \to J/\psi+ X$ process
\cite{DELPH}. Similarly, the recent experimental data on the $J/\psi$
photoproduction of H1 \cite{H1} are fairly well described by the
complete NLO NRQCD corrections \cite{kniehl1}, which give strong
support to the existence of the COM. However, at B-factories, a
series of processes were calculated up to the QCD NLO corrections in
the CSM \cite{zhang,wang,sang}. Together with the relativistic
correction \cite{RC,RC2}, it seems that most experimental data could
be understood. Additionally, the $J/\psi$ polarization in
hadroproduction at the Tevatron \cite{nnrtev} and photoproduction at
the HERA \cite{nnrHERA} also conflict with the NRQCD predictions. So
the existence of the COM is still under question and far from being
proven. Therefore, the further test for the CSM and COM under NRQCD
in heavy quarkonium production is still needed.

\par
In order to investigate the effects of the CSM and COM in heavy
quarkonium physics, more processes of heavy quarkonium production
and decay should be studied. At the LHC, with more attention for the
heavy quarkonium production, much work has been done in these
regions \cite{quLHC,quLHC2}, and some channels have been calculated
to the NLO \cite{quNLO,liw,songw}. As quarkonium can be identified
by using their purely leptonic decays \cite{bager:111,
braaten:091501}, the bottom quark can be identified by
reconstructing secondary vertices, and the high-$p_T$ bottom quark
can be tagged with reasonably high efficiency at the LHC, meanwhile
the observation of a bottom quark with high-$p_T$ can reduce the
backgrounds of the heavy quarkonium production. Thus, it is also very
interesting to study the prompt heavy quarkonium production
associated with a massive (anti)bottom quark at the LHC. These
processes have the potential to be detected and can provide an
interesting signature at the LHC. The measurement of these processes
is useful to test the CSM and COM.

\par
In this work, we perform the calculations for prompt heavy
quarkonium production in association with a massive (anti)bottom
quark at the LHC. Within this work, the mass of bottom quark is
retained in all the partonic processes. The paper is organized as
follows: We present the details of the calculation strategies in
Sec.II. The numerical results are given in Sec.III.
Finally, a short summary and discussions are given.

\vskip 5mm
\section{The details of the calculation}\label{calc}
The purpose of this paper is to study the associated production of
prompt heavy quarkonium and a massive (anti)bottom quark to LO in
the NRQCD factorization formalism at the LHC. We denote the heavy
quarkonium as $\mathcal {Q}$.  As we know, the cross section for the
$g(p_{1})+b(p_{2})\to \mathcal {Q}(k_{3})+b(k_{4})$ partonic process
in the SM should be the same as that for its charge conjugate subprocess
$g(p_{1})+\bar{b}(p_{2})\to \mathcal {Q}(k_{3})+\bar{b}(k_{4})$, and
the luminosity of the bottom quark in a proton is same as that of the
antibottom quark. Therefore, the production rates of the
$\mathcal {Q}b$ and the $\mathcal {Q}\bar b$ are
identical at the LHC. In the following sections, we present only the
analytical calculations of the related partonic process
$g(p_{1})+b(p_{2})\to \mathcal {Q}(k_{3})+b(k_{4})$ and the parent
process $pp\to \mathcal {Q}+b$ unless otherwise indicated.

\par
The cross section for the $pp\to \mathcal {Q}+b$ process is
expressed as
\begin{eqnarray}
\sigma\left(pp \to \mathcal {Q}+b\right)&=& \int
dx_1dx_2\sum_{n}\langle{\cal O}^{\mathcal
{Q}}[n]\rangle\hat{\sigma}\left(gb\to
Q\bar{Q}[n]+b\right)\nb \\
&&\times \left[G_{g/A}(x_1,\mu_f)G_{b/B}(x_2,\mu_f)+(A
\leftrightarrow B )\right].
\end{eqnarray}
Here $\hat{\sigma}\left(gb\to Q\bar{Q}[n]+b\right)$ describes the
short-distance production of a heavy $Q\bar{Q}$ pair in the color,
spin and angular momentum state $n$. When $Q=c$, $\mathcal {Q}$
represents charmonium and when $Q=b$, $\mathcal {Q}$ is bottomonium.
$\langle{\cal O}^{\mathcal {Q}}[n]\rangle$ is the long-distance
matrix element, which describes the hadronization of the heavy
$Q\bar{Q}$ pair into the observable quarkonium state $ \mathcal
{Q}$. $G_{b,g/A,B}$ are the parton distribution functions. A
and B refer to protons at the LHC. The indices $g,b$ represents the
gluon and bottom quark, respectively.

The short-distance cross section for the production of a $Q\bar{Q}$
pair in a Fock state $n$, $\hat{\sigma}[gb\to Q\bar{Q}[n]+b]$, is
calculated from the amplitudes which are obtained by applying
certain projectors onto the usual QCD amplitudes for open $Q\bar{Q}$
production. In the notations of Ref. \cite{p2}:
\begin{eqnarray}
{\cal A}_{Q\bar{Q} [{}^1S_0^{(1/8)} ]} = {\rm Tr} \Big[ {\cal
C}_{1/8} \Pi_0 {\cal A} \Big]_{q=0}, \nonumber
\end{eqnarray}
\begin{eqnarray}
{\cal A}_{Q\bar{Q} [ ^3S_1^{(1/8)} ]} = {\cal E}_{\alpha} {\rm Tr}
\Big[ {\cal C}_{1/8} \Pi_{1}^{\alpha} {\cal A} \Big]_{q=0},
\nonumber
\end{eqnarray}
\begin{eqnarray}
{\cal A}_{Q\bar{Q} [ ^1P_1^{(1/8)} ]} = {\cal E}_{\alpha }
\frac{d}{dq_{\alpha}} {\rm Tr} \Big[ {\cal C}_{1/8} \Pi_0 {\cal A}
\Big]_{q=0}, \nonumber
\end{eqnarray}
\begin{eqnarray}
{\cal A}_{Q\bar{Q} [ ^3P_J^{(1/8)} ]} = {\cal E}_{\alpha
\beta}^{(J)} \frac{d}{dq_{\beta}} {\rm Tr} \Big[ {\cal C}_{(1/8)}
\Pi_1^{\alpha} {\cal A} \Big]_{q=0},
\end{eqnarray}
where ${\cal A}$ denotes the QCD amplitude with amputated heavy-quark spinors,
the lower index $q$ represents the momentum of the
heavy-quark in the $Q\bar{Q}$ rest frame. $\Pi_{0/1}$ are spin
projectors onto the spin singlet and spin triplet states. ${\cal
C}_{1/8}$ are color projectors onto the color-singlet and color-octet
states. ${\cal E}_{\alpha}$ and ${\cal E}_{\alpha \beta}$
represent the polarization vector and tensor of the $Q\bar{Q}$
states, respectively.

\par
Then the LO short-distance cross section for the partonic process
$g(p_{1})+b(p_{2})\to Q\bar{Q}[n](k_{3})+b(k_{4})$ is obtained by
using the following formula:
\begin{eqnarray}
\hat{\sigma}\left(gb\to Q\bar{Q}[n]+b\right)=\frac{1}{16\pi
\hat{s}^2 N_{col}N_{pol}} \int_{\hat{t}_{min}}^{\hat{t}_{max}}
d\hat{t}\overline{\sum}|{\cal A}_{Q\bar{Q} [n]}|^2.
\end{eqnarray}
The summation is taken over the spins and colors of initial and
final states, and the bar over the summation denotes averaging over
the spins and colors of initial partons. The Mandelstam variables
are defined as:
$\hat{s}=(p_1+p_2)^2,\hat{t}=(p_1-k_3)^2,\hat{u}=(p_1-k_4)^2$.
$N_{col}$ and $N_{pol}$ refer to the numbers of colors and polarization of
states $n$, separately \cite{p2}. In
this paper, we present all contributing partonic cross sections in
analytic form in the Appendix.
\begin{figure}[!htb]
\begin{center}
\begin{tabular}{cc}
{\includegraphics[width=10cm]{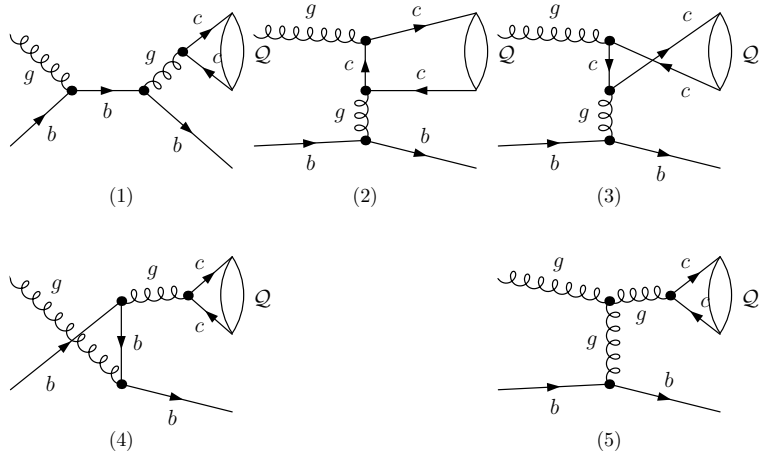}}
\end{tabular}
\end{center}
 \vspace*{-0.7cm}\caption{The LO Feynman diagrams  for prompt charmonium
production in association with a massive bottom quark at the LHC.}
\label{f1}
\end{figure}

\begin{figure}[!htb]
\begin{center}
\begin{tabular}{cc}
{\includegraphics[width=10cm]{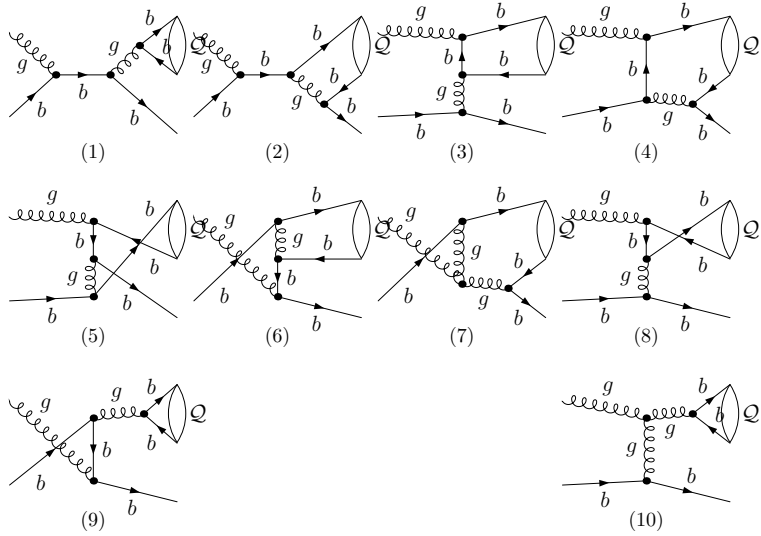}}
\end{tabular}
\end{center}
\vspace*{-0.7cm} \caption{The LO Feynman diagrams  for prompt
bottomonium production in association with a massive bottom quark at
the LHC.} \label{f2}
\end{figure}
In the case of prompt charmonium production in association with a massive
bottom quark, there are five Feynman diagrams that contribute to this
process at LO; we present them in Fig.\ref{f1}. There are 10 Feynman
diagrams for prompt bottomonium partonic process, which are drawn in Fig.\ref{f2}. In our
calculations, the $Q\bar{Q}$ Fock states contributing at LO in $v$ for
$\mathcal {Q}$ are shown in Table \ref{f3}.

\begin{table}[!htb]
\vspace*{-3cm}
\begin{center}
\begin{tabular}{cc}
{\includegraphics[width=17cm]{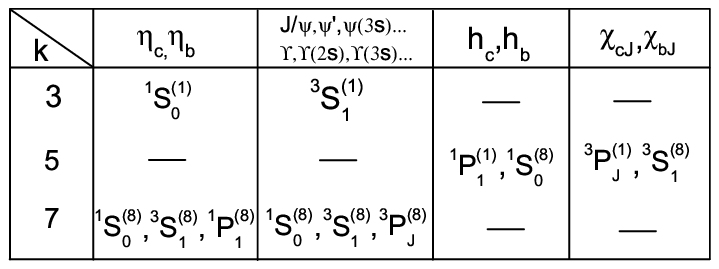}}
\end{tabular}
\end{center}
\vspace*{-6cm}
 \caption{Values of $k$ in the velocity-scaling rule
$\langle{\cal O}^{\mathcal {Q}}[n]\rangle \propto v^k$ for the
leading $Q\bar{Q}$ Fock states $n$ pertinent to $\mathcal {Q}$.}
\label{f3}
\end{table}

Following the heavy-quark spin symmetry, these multiplicity
relations of LDMEs
\begin{eqnarray}
<{\cal O}^{\psi(nS)}[{}^3\!P_J^{(8)}]> &=&(2J+1) <{\cal
O}^{\psi(nS)}[{}^3\!P_0^{(8)}]>,
\nonumber\\
<{\cal O}^{\Upsilon(nS)}[{}^3\!P_J^{(8)}]> &=&(2J+1) <{\cal
O}^{\Upsilon(nS)}[{}^3\!P_0^{(8)}]>,
\nonumber\\
<{\cal O}^{\chi_{QJ}}[{}^3\!P_J^{(1)}]> &=&(2J+1) <{\cal
O}^{\chi_{Q0}}[{}^3\!P_0^{(1)}]>,
\nonumber\\
<{\cal O}^{\chi_{QJ}}[{}^3\!S_1^{(8)}]> &=&(2J+1) <{\cal
O}^{\chi_{Q0}}[{}^3\!S_1^{(8)}]> \label{eq:mul}
\end{eqnarray}
be assumed satisfied \cite{bbl}.

\section{Numerical results}

In our numerical calculations, we focus on the cases $\mathcal
{Q}=J/\psi,\chi_{cJ},\Upsilon,\chi_{bJ}$. These quarkoniums can be
efficiently identified experimentally, and their LDMEs are relatively
well constrained \cite{Yan-Qing
Ma:2010,Butenschoen:2010px,Eichten:1995PRD,u1m,u8m}. The $\eta_{c},
\eta_{b}, h_c$ and $h_b$ mesons are more difficult to detect
experimentally, and we will give the differential cross sections
relevant to these quarkoniums analytically in our Appendix.

We take CTEQ6L1 parton distribution functions \cite{CTEQ6} with a one-loop running
$\alpha_s$ in the LO calculations, and the corresponding fitted
value $\alpha_s(M_Z) = 0.130$ is used for our calculations. The
factorization scale is chosen as $\mu_f = m_T$, where $m_T =
\sqrt{\Big( p_T^{\mathcal {Q}} \Big)^2 + m_{\mathcal {Q}}^2}$ is the
$\mathcal {Q}$ transverse mass. The masses of heavy quark are set as
$m_c = 1.5~{\rm GeV}$ and $m_b =4.75~{\rm
 GeV}$.

\par
For CO LDME of $J/\psi$, it can be extracted from experimental data or from lattice QCD
calculations. Before lattice QCD giving out the results, the
color-octet matrix elements are determined only by fitting the
theoretical prediction to experimental data. As done in the
literature \cite{RC2,Butenschoen:2010px}, the color-octet matrix
elements are extracted from experimental data of $J/\psi$ production
in $e^+e^-$, $ep$ collisions and hadron collisions. In Ref.
\cite{RC2}, through the analysis of the process $e^+e^- \to
J/\psi + X_{non-c\bar{c}}$ at B factories, they give a very
stringent constraint on the CO contribution, and imply that the
values of color-octet matrix elements are very small, but their
results may not be consistent with the naive velocity-scaling rules.
Recently, M. Butenschoen and B.A. Kniehl performed a multiprocess
fit of the CO LDMEs \cite{Butenschoen:2010px}, give more global
fit results, and their results are consistent with NRQCD scaling
rules. In our work, though we performed a LO calculation, we will
tentatively choose the color-octet matrix elements obtained by the
M. Butenschoen and B.A. Kniehl's NLO fit results \cite{Butenschoen:2010px}
\begin{eqnarray}
< O^{J/\psi}[{}^1S_0^{(8)}]>&=& (4.76\pm0.71)\times10^{-2}  GeV^3, \nb \\
< O^{J/\psi}[{}^3S_1^{(8)}]> &=& (0.265\pm0.91)\times10^{-2}  GeV^3, \nb \\
< O^{J/\psi}[{}^3P_0^{(8)}]> &=& (-1.32\pm0.35)\times10^{-2} GeV^5,
\end{eqnarray} as our input parameters.
The relation of the CS matrix elements $<
O^{\chi_{c0}}[{}^3P_0^{(1)}]> $
 of $\chi_{c0}$ with the P-wave function at the origin is the
 formula $< O^{\chi_{c0}}[{}^3P_0^{(1)}]>
=\frac{3N_c}{2\pi}|R_P^{\prime}(0)|^2$,  and we choose
$|R_P^{\prime}(0)|^2=0.075 GeV^{5}$ from the potential model
calculations \cite{Eichten:1995PRD}. For the CO matrix element $<
O^{\chi_{c0}}[{}^3S_1^{(8)}]> $,  we use $<
O^{\chi_{c0}}[{}^3S_1^{(8)}]> \approx 2.2\times10^{-3} GeV^3$ as our
input parameter \cite{Yan-Qing Ma:2010}. For the NRQCD matrix
elements of bottomonium, the CS matrix elements are taken from the
potential model calculations of \cite{u1m}, and the CO matrix
elements are determined from the CDF data \cite{u8m}. In our
calculations, the relation of CS matrix elements with the
conventions  matrix elements of Bodwin-Braaten-Lepage is always considered carefully
\cite{p2}.

In our numerical calculations, we summed the contributions of $pp\to \mathcal {Q}+b$ and $pp\to
\mathcal {Q}+\bar{b}$. If $\mathcal {Q}=\chi_{QJ}$, we also summed
the contributions of $\chi_{Q0},\chi_{Q1}$ and $\chi_{Q2}$. We take
the constraints of $|y^{\mathcal {Q}}| < 3$ for heavy quarkoniums,
and $p_{T,b(\bar{b})}
> 5~{\rm GeV}$, $|y_{b(\bar{b})}| < 3$ for $b(\bar{b})$ quark. The colliding
energy used in this paper is 14 $TeV$.
\begin{figure}[!htb]
\vspace*{-1cm}
\begin{center}
\begin{tabular}{cc}
{\includegraphics[width=8.6cm]{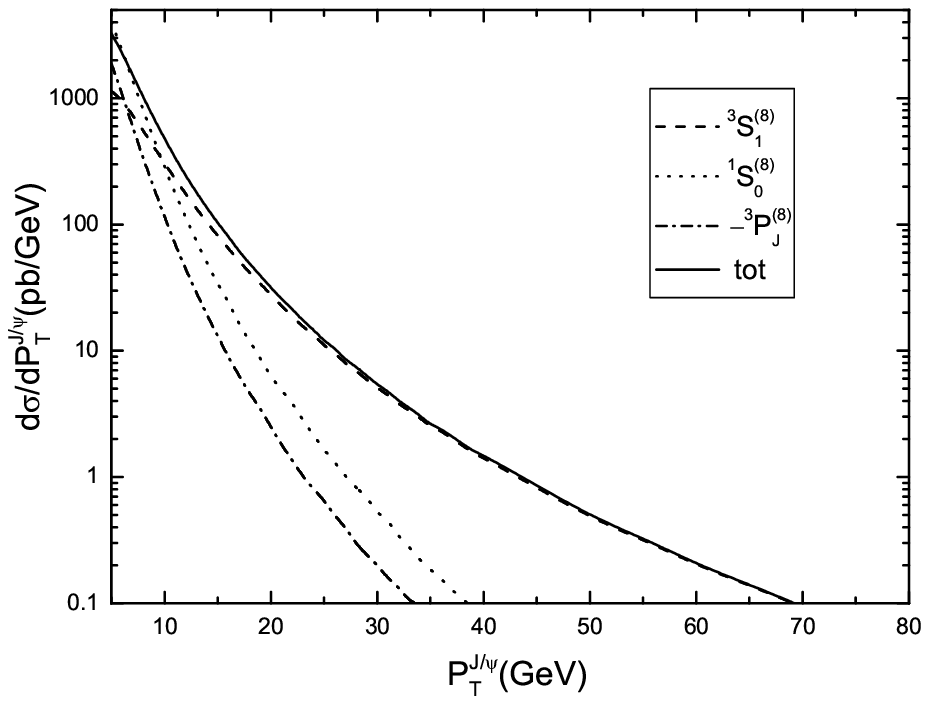}}
{\includegraphics[width=8.6cm]{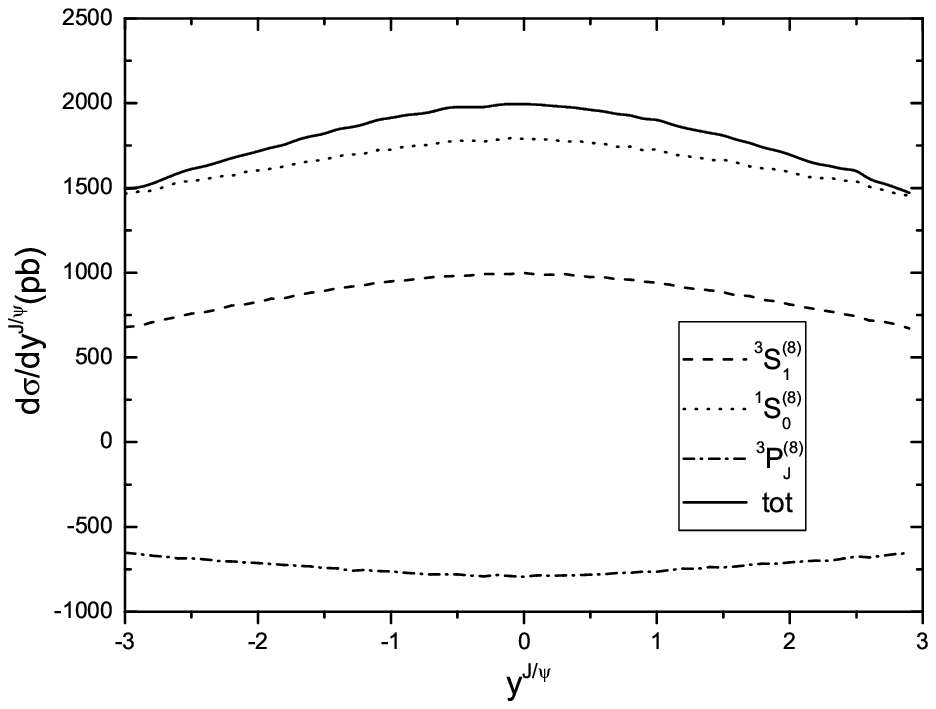}}
\end{tabular}
\end{center}
\vspace*{-1cm}
 \caption{The LO distributions of $p_T^{J/\psi}$ and $y^{J/\psi}$ for the $pp \to
J/\psi+b(\bar{b})$ process, and the contributions of the $c\bar{c} [
^1S_0^{(8)} ]$ , $c\bar{c} [ ^3S_1^{(8)}]$ and $c\bar{c}[
^3P_J^{(8)} ]$ Fock states at the LHC.} \label{jpt}
\end{figure}

\par
In Fig.\ref{jpt}, we present the LO distributions of $p_T^{J/\psi}$
and $y^{J/\psi}$ for the process $pp\to J/\psi+b(\bar{b})$ at the
LHC. At LO, the associated production of prompt $J/\psi$ with a
(anti)bottom quark is forbidden in the CSM, and there is only CO
contribution in this process up to the $\alpha^3_s v^7$ order within the NRQCD framework. For
comparison, we also depicted the contributions of the $c\bar{c} [
^1S_0^{(8)} ]$, $c\bar{c} [ ^3S_1^{(8)} ]$ and $c\bar{c} [
^3P_J^{(8)}]$ Fock states in these figures. For the LDME of $
c\bar{c} [ ^3P_J^{(8)}]$ ($J=0,1,2$) Fock states are negative, the
contribution of these Fock states are negative too. In
Fig.\ref{jpt}, we present the $p_T$ distributions
$-d\sigma/dp_T^{J/\psi}[c\bar{c}[ ^3P_J^{(8)} ]]$ in the $p_T$ distribution figure. From
the figure we can see that when $p_T$ is smaller than about 10 GeV,
the ${}^1S_0^{(8)}$ state gives the main contribution to the $p_T$
distribution of prompt $J/\psi$ with a (anti)bottom quark
production. With the $p_T$ of $J/\psi$ increase, the contributions
of $c\bar{c} [ ^1S_0^{(8)} ]$ and $ c\bar{c} [ ^3P_J^{(8)} ]$ Fock
states quickly decrease. The differential cross section is
dominated by the $c\bar{c} [ ^3S_1^{(8)}]$ Fock-state contribution
at large $p_T$ region. In the range of $5~{\rm GeV} < p_T^{J/\psi} <
50~{\rm GeV}$, the ${d\sigma/dp_T^{J/\psi}}$ is in the range of
$[0.486,~3276.396]pb/GeV$, and it reaches the maximum when
$p_T^{J/\psi} = 5~{\rm GeV}$.

\begin{figure}[!htb]
\vspace*{-1cm}
\begin{center}
\begin{tabular}{cc}
{\includegraphics[width=8.6cm]{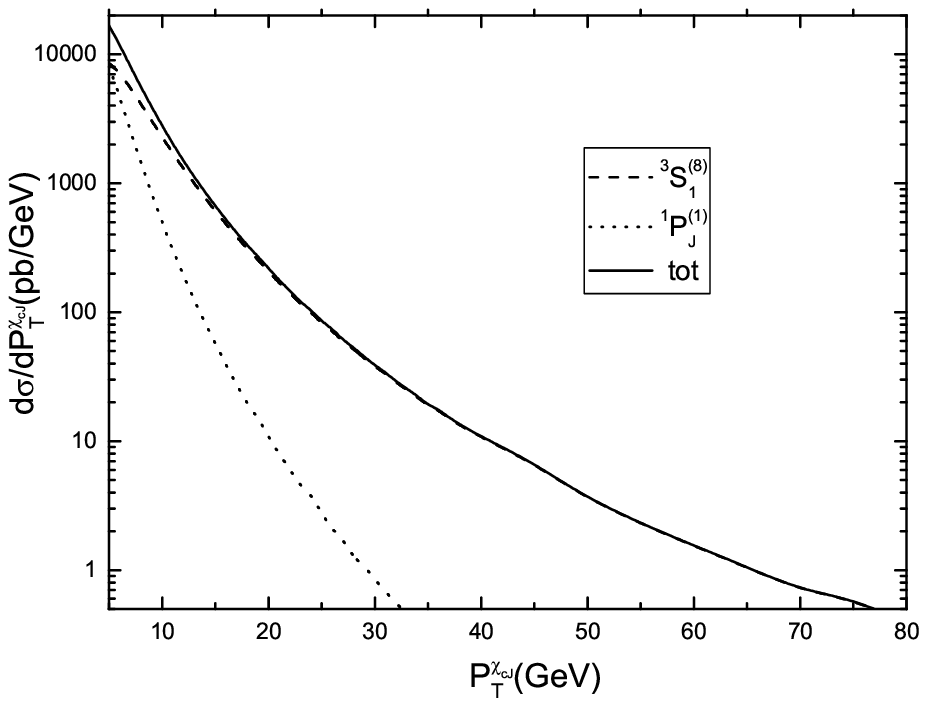}}
{\includegraphics[width=8.6cm]{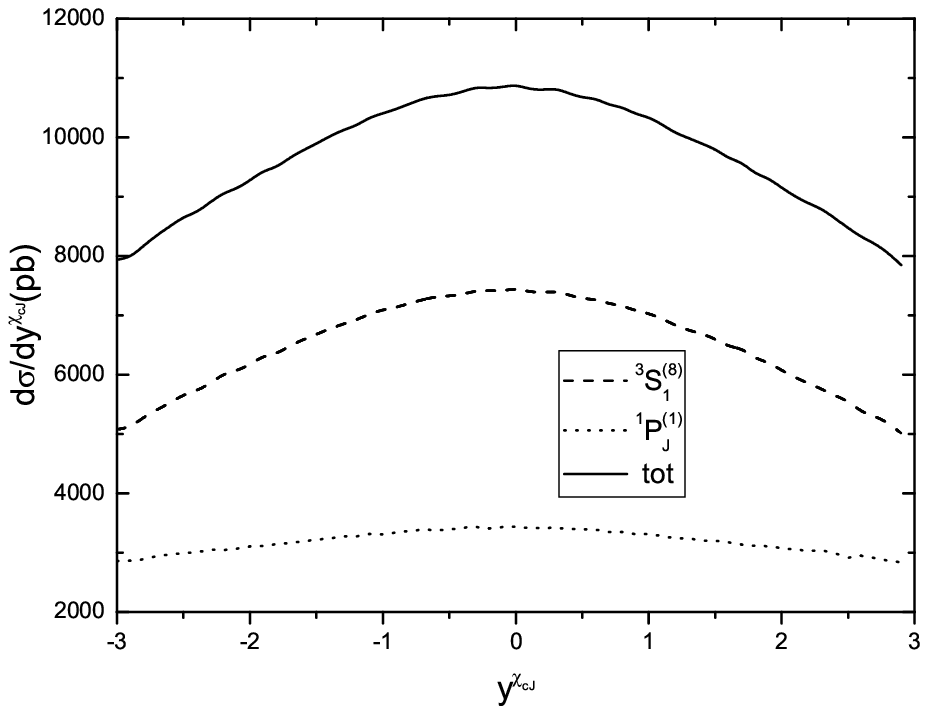}}
\end{tabular}
\end{center}
 \vspace*{-1cm}\caption{The LO distributions of $p_T^{\chi_{cJ}}$ and $y^{\chi_{cJ}}$ for the $pp \to \chi_{cJ}+b(\bar{b})$
process, and the contributions of the $c\bar{c} [ ^3S_1^{(8)} ]$ and
$c\bar{c}[ ^3P_J^{(1)}]$ Fock states at the LHC.} \label{kcpt}
\end{figure}

The curves for the LO distributions of $p_T^{\chi_{cJ}}$ and
$y^{\chi_{cJ}}$ for the process $pp\to \chi_{cJ}+b(\bar{b})$ at the
LHC are drawn in Fig.\ref{kcpt}. From the figure we can see that the
contribution from CS is about the same order of magnitude with CO at
$p_T\simeq 5$ GeV. The CO contribution dominates over production at
the large $p_T$ region, and it decreases much more slowly than that of CS
as $p_T$ increases. From Fig.\ref{jpt} and Fig.\ref{kcpt}, we can
find that the cross section of $J/\psi+b(\bar{b})$ and
$\chi_{cJ}+b(\bar{b})$ associated production mainly comes from the
$c\bar{c} [ ^3S_1^{(8)} ]$ Fock state contribution at large $p_T$
region. The cross section ratios of $J/\psi+b(\bar{b})$ and
$\chi_{cJ}+b(\bar{b})$ associated production can be estimated
approximately by $<{\cal O}^{J/\psi}[{}^3\!S_1^{(8)}]> /
\sum_{J=0}^2 <{\cal O}^{\chi_{cJ}}[{}^3\!S_1^{(8)}]>$
${}\approx0.134$  at large $p_T$ region \cite{Yan-Qing
Ma:2010,Butenschoen:2010px}. So when we consider the process prompt
$J/\psi$ with a (anti)bottom quark associated production at the LHC,
the indirect prompt production for this process via radiative decays
of $\chi_{cJ}$ should be included.

\begin{figure}[!htb]
\vspace*{-1cm}
\begin{center}
\begin{tabular}{cc}
{\includegraphics[width=8.6cm]{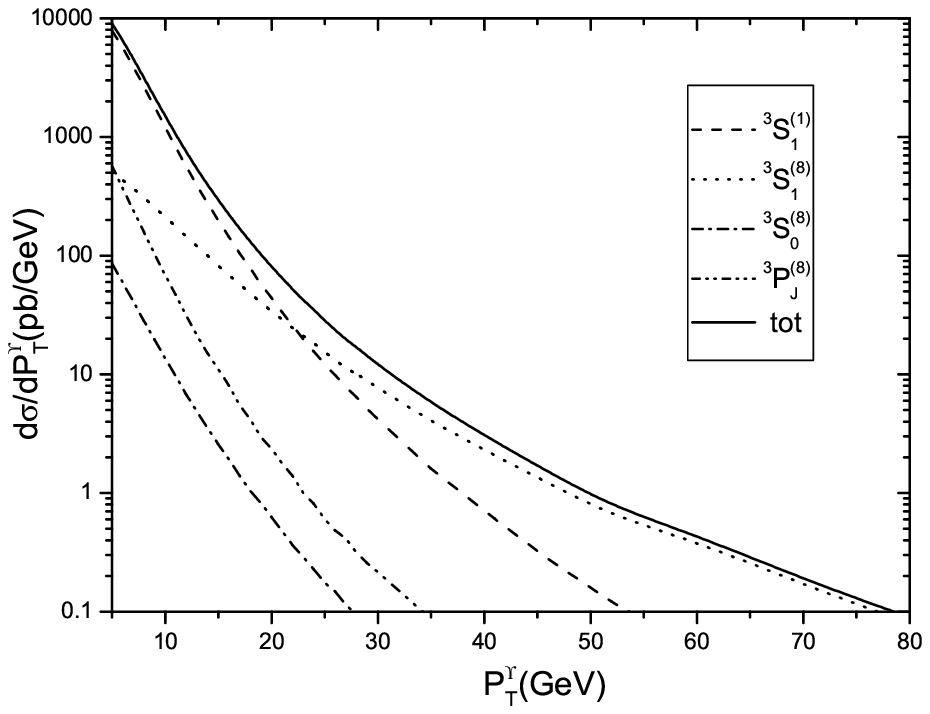}}
{\includegraphics[width=8.6cm]{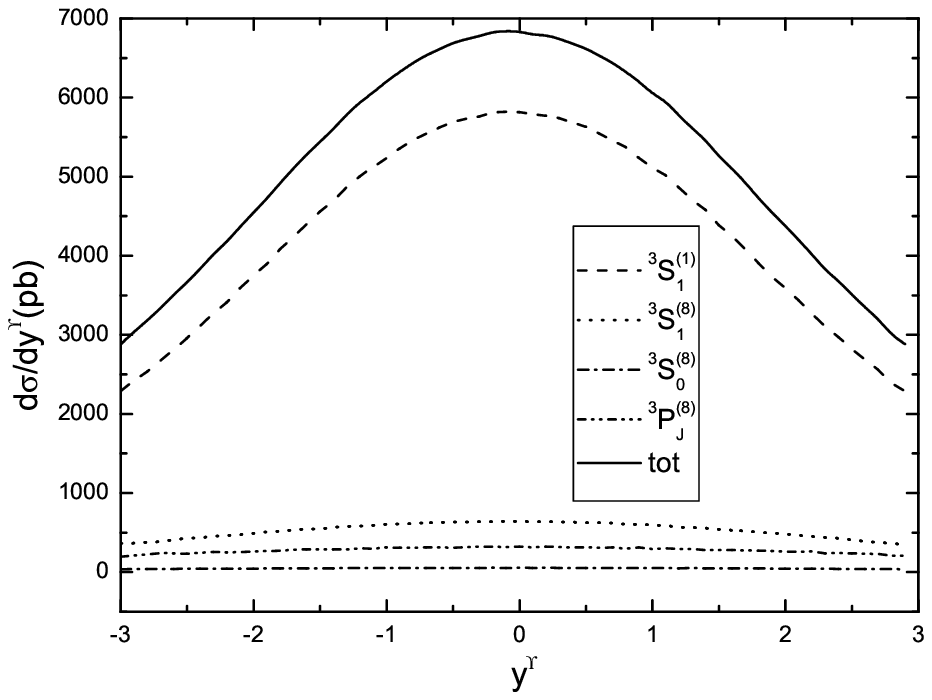}}
\end{tabular}
\end{center}
 \vspace*{-1cm}
 \caption{The LO distributions of $p_T^{\Upsilon}$ and $y^{\Upsilon}$ for the process
$pp \to \Upsilon+b(\bar{b})$, and the contributions of the $b\bar{b} [ ^3S_1^{(1)}
]$, $b\bar{b} [ ^1S_0^{(8)}]$, $b\bar{b} [ ^3S_1^{(8)} ]$ and
$b\bar{b}[ ^3P_J^{(8)}]$ Fock states at the LHC.} \label{upt}
\end{figure}

The LO distributions of $p_T^{\Upsilon}$ and $y^{\Upsilon}$ for the process
$pp \to \Upsilon+b(\bar{b})$, and the contributions of the
$b\bar{b} [ ^3S_1^{(1)} ]$, $b\bar{b} [ ^1S_0^{(8)} ]$, $b\bar{b} [
^3S_1^{(8)} ]$ and $b\bar{b}[ ^3P_J^{(8)}]$ Fock states at the LHC
are illustrated in Fig.\ref{upt}. We can see that the differential
cross section of $b\bar{b} [ ^3S_1^{(1)} ]$ and $b\bar{b} [
^3S_1^{(8)} ]$ Fock states give the main contribution to the
distributions of $p_T^{\Upsilon}$ for the process $pp \to
\Upsilon+b(\bar{b})$ at the LO. The $b\bar{b} [ ^3S_1^{(1)}
]$ state give the main contribution to the $p_T$ distribution at low
$p_T$ region. With the increase of $p_T^\Upsilon$, the
contribution of $b\bar{b} [ ^3S_1^{(1)} ]$ Fock state quickly
decrease and the contribution of $b\bar{b} [ ^3S_1^{(8)} ]$  Fock
state becomes more important. The contribution of $b\bar{b} [
^3S_1^{(8)} ]$  Fock states starts to be comparable to $b\bar{b} [
^3S_1^{(1)} ]$ contribution for $p_T\simeq 22$ GeV. The differential
cross section is dominated by the $b\bar{b} [ ^3S_1^{(8)}]$ Fock
state contribution at large $p_T$ region.

In this paper, we have only considered the LO process $pp\to
\Upsilon+b(\bar{b})+X$ at $\alpha^3_s$. In Ref. \cite{quLHC2}, the
authors have evaluated the color-singlet contribution to the process $pp\to
\Upsilon+b\bar{b}$. One of the properties of $pp\to
\Upsilon+b\bar{b}$ is that its CS contribution (which is at
$\alpha^4_s$) already includes $p^{-4}_T$ topologies which dominate
at large $p_T$. If we consider the process $pp\to
\Upsilon+b(\bar{b})+X$ at $\alpha^4_s$, the process $pp\to
\Upsilon+b\bar{b}$ may give important CS contribution in large $p_T$
region.

\begin{figure}[!htb]
 \vspace*{-1cm}
\begin{center}
\begin{tabular}{cc}
{\includegraphics[width=8.6cm]{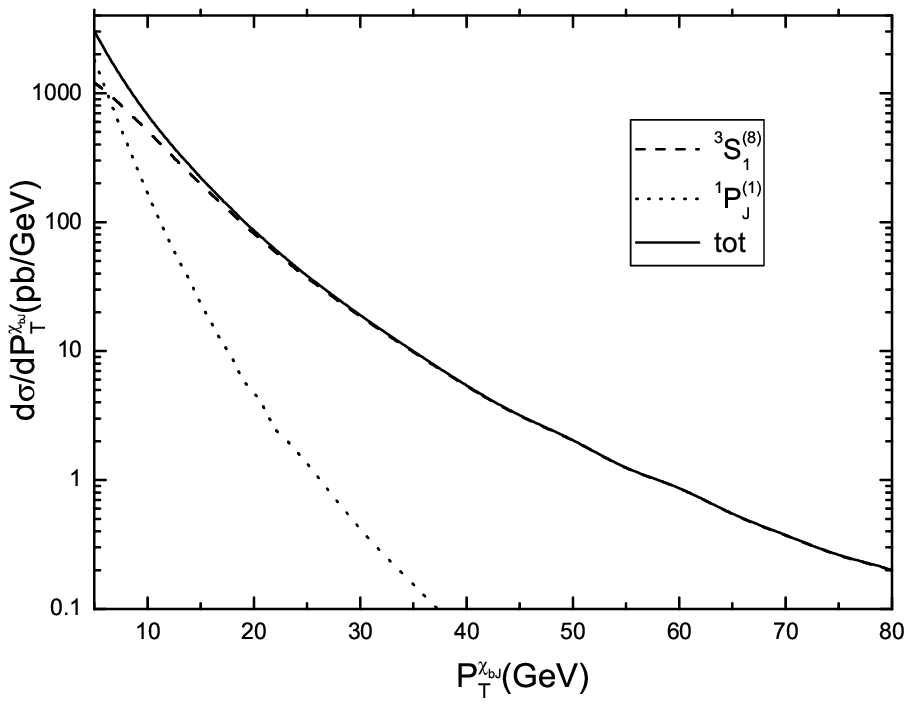}}
{\includegraphics[width=8.6cm]{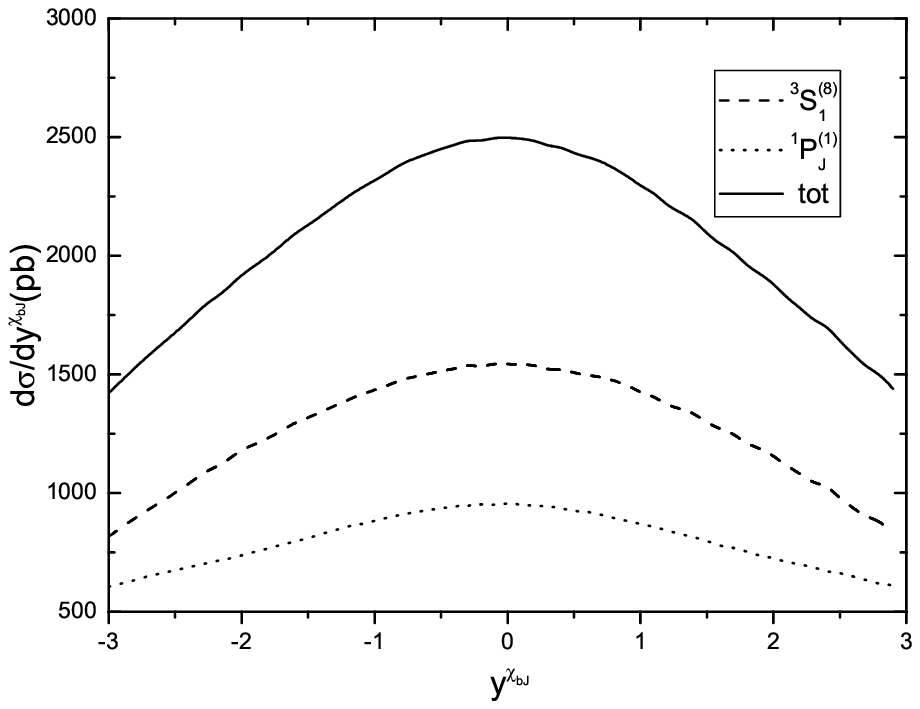}}
\end{tabular}
\end{center}
 \vspace*{-1cm}
 \caption{The LO distributions of $p_T^{\chi_{bJ}}$ and $y^{\chi_{bJ}}$ for the $pp \to \chi_{bJ}+b(\bar{b})$
process, and the contributions of the $b\bar{b} [ ^3S_1^{(8)} ]$ and
$b\bar{b}[ ^3P_J^{(1)}]$ Fock states at the LHC.} \label{kbpt}
\end{figure}

Finally, we give the LO distributions of $p_T^{\chi_{bJ}}$ and
$y^{\chi_{bJ}}$ for the process $pp \to \chi_{bJ}+b(\bar{b})$, and
the contributions of the $b\bar{b}[ ^3S_1^{(8)} ]$ and $b\bar{b}[
^3P_J^{(1)}]$ Fock states at the LHC in Fig.\ref{kbpt}. From these
figures we can see that the contribution from CS is larger than CO when
$p_T < 6$ GeV. The CO contribution dominates over production at the
large $p_T$ region, and it decreases much slower than that of CS as
$p_T$ increases. In the range of $5~{\rm GeV} < p_T^{\chi_{bJ}} <
50~{\rm GeV}$, the ${d\sigma/dp_T^{\chi_{bJ}}}$is in the range of
$[2.095,~3022.305]pb/GeV$, and it reaches the maximum when
$p_T^{J/\psi} = 5~{\rm GeV}$.

\section{Discussion and summary }
In this paper, we investigate the associated production of prompt
heavy quarkonium with a massive (anti)bottom quark to LO in the
NRQCD factorization formalism at the LHC. We have considered all
experimentally established heavy quarkoniums, with
${}^{2S+1}L_J={}^1\!S_0,{}^3\!S_1,{}^1\!P_1,{}^3\!P_J$, and listed the
differential cross sections relevant to these quarkonium
analytically in our Appendix. We present the numerical predictions
of the differential cross sections of the $p_T^{\mathcal{Q}}$ and
rapidity $y^{\mathcal{Q}}$ for $\mathcal{Q}=J/\psi,\chi_{cJ},\Upsilon,\chi_{bJ}$ at the LHC. We
find that the associated production of prompt
$\mathcal{Q}=J/\psi,\chi_{cJ},\Upsilon,\chi_{bJ}$ and a massive
(anti)bottom quark at the LHC have the potential to be detected.
When $p_T$ is smaller than about 10 GeV, the $c\bar{c} [ ^1S_0^{(8)}
]$ state give the main contribution to the $p_T$ distribution of
prompt $J/\psi$ with a massive (anti)bottom quark production. For the
process $pp \to \Upsilon+b(\bar{b})$, the contribution of the CSM
is larger than that in the COM at low $p_T$ region. We also
investigate the processes of $pp\to \chi_{cJ}+b(\bar{b})$ and $pp
\to \chi_{bJ}+b(\bar{b})$, the $p_T$ distribution are dominated by
the CO Fock state contribution at the large $p_T$ region. These
processes provide an interesting signature that could be studied
at the LHC, and the measurement of these processes is useful to test the CSM and COM.

\vskip 10mm
\par
\noindent{\large\bf Acknowledgments:} This work was supported in
part by  the Key Research Foundation of Education Ministry of Anhui
Province of China (NO.KJ2012A021), the Youth Foundation of Anhui University,
and financed by the 211 Project of Anhui University.

\begin{appendix}
\setcounter{equation}{0}
\section*{Appendix}

In this Appendix, we list the differential cross sections
$d\hat{\sigma}/d\hat{t}$ for processes $gb\to QQ[n]+b$. Our results
read

\par
\begin{eqnarray*}
\frac{d\sigma}{d\hat{t}}\left(gb\to c\overline{c}
\left[{}^1\!S_0^{(1)}\right]b\right)=&&
\frac{4 \alpha^3_{s} \pi^2}{9 m_{c} \hat{t}^2 (-4 m^2_{c} + \hat{t})^2 \hat{s}^2} {[2 m^4_{b} (8 m^2_{c} - \hat{t}) - 16 m^4_{c} (\hat{s} + \hat{u}) + 4 m^2_{c} (\hat{s}} \\
&&{+ \hat{u}) (\hat{s} + \hat{t} + \hat{u})
- \hat{t} (\hat{s}^2 + \hat{u}^2) - 2 m^2 _{b} (\hat{t} (-\hat{s} + \hat{t} - \hat{u}) + m^2_{c} (8 \hat{s} }  \\
&&{ - 4 \hat{t} + 8 \hat{u}))]} ,
\end{eqnarray*}
\par
\begin{eqnarray*}
\frac{d\sigma}{d\hat{t}}\left(gb\to c\overline{c}
\left[{}^1\!S_0^{(8)}\right]b\right)=&& \frac {5 \alpha^3_{s}
\pi^2}{36 m_{c} \hat{t}^2 (-4 m^2_{c} + \hat{t})^2 \hat{s}^2} {[2
m^4_{b} (8 m^2_{c} - \hat{t}) - 16 m^4_{c} (\hat{s} + \hat{u}) +
   4 m^2_{c} (\hat{s} + \hat{u}) (\hat{s} } \\
&&{ + \hat{t} + \hat{u}) - \hat{t} (\hat{s}^2 + \hat{u}^2) - 2
m^2_{b} (\hat{t} (-\hat{s} + \hat{t} - \hat{u}) + m^2_{c} (8 \hat{s}
- 4 \hat{t} + 8 \hat{u}))]} ,
\end{eqnarray*}

\par
\begin{eqnarray*}
\frac{d\sigma}{d\hat{t}}\left(gb\to c\overline{c}
\left[{}^3\!S_1^{(1)}\right]b\right) = 0 ,
\end{eqnarray*}

\par
\begin{eqnarray*}
\frac{d\sigma}{d\hat{t}}\left(gb\to c\overline{c}
\left[{}^3\!S_1^{(8)}\right]b\right)=&& \frac {\alpha^3_{s} \pi^2 (7
m^4_{b} + 4 \hat{s}^2 - \hat{s} \hat{u} + 4 \hat{u}^2 - 7 m^2_{b}
(\hat{s} + \hat{u}))}{216
 m^3_{c} (m^2_{b} - \hat{s})^2 (m^2_{b} - \hat{u})^2 (-2 m^2_{b} + \hat{s} + \hat{u})^2 \hat{s}^2}
{ [6 m^8_{b} - \hat{s} \hat{u} (32 m^4_{c} } \\
&&{
+ \hat{s}^2 + \hat{u}^2 - 8 m^2 _{c} (\hat{s} + \hat{u})) - m^4_{b} (32 m^4_{c} + 3 \hat{s}^2 + 14 \hat{s} \hat{u} + 3 \hat{u}^2 - 8 m^2_{c} } \\
&&{
 (\hat{s} + \hat{u})) + m^2_{b} (\hat{s}^3 - 32 m^2_{c} \hat{s} \hat{u} + 7 \hat{s}^2 \hat{u} + 7 \hat{s} \hat{u}^2 + \hat{u}^3 + 32 m^4_{c} (\hat{s} + \hat{u}))]} ,
\end{eqnarray*}
\par
\begin{eqnarray*}
\frac{d\sigma}{d\hat{t}}\left(gb\to c\overline{c}
\left[{}^1\!P_1^{(1)}\right]b\right) = 0 ,
\end{eqnarray*}

\par
\begin{eqnarray*}
\frac{d\sigma}{d\hat{t}}\left(gb\to c\overline{c}
\left[{}^1\!P_1^{(8)}\right]b\right)=&& \frac{-\alpha^3_{s}
\pi^2}{12 m^3_{c} \hat{t}^2 (-4 m^2_{c} + \hat{t})^2 \hat{s}^2}
{ [2 m^4_{b} \hat{t} + 4 m^2_{b} (8 m^4_{c} - 2 m^2_{c} \hat{t} - \hat{s} \hat{t}) }  \\
&&{+ \hat{t} (16 m^4_{c} + 2 \hat{s}^2 + 2 \hat{s} \hat{t} +
\hat{t}^2 - 8 m^2_{c} (\hat{s} + \hat{t}))]} ,
\end{eqnarray*}

\par
\begin{eqnarray*}
\frac{d\sigma}{d\hat{t}}\left(gb\to c\overline{c}
\left[{}^3\!P_J^{(1)}\right]b\right)=&& \frac{4 \alpha^3_{s}
\pi^2}{9 \hat{t}^2 (-4 m^3_{c} + m_{c} \hat{t})^3 \hat{s}^2}
{[2 m^4_{b} (28 m^2_{c} - 3 \hat{t}) \hat{t} + 4 m^2_{b} (224 m^6_{c} - 16 m^4_{c} \hat{t} }  \\
&&{ + 3 s \hat{t}^2 - 2 m^2_{c} \hat{t} (14 \hat{s} + 5 \hat{t})) +
\hat{t} (448 m^6_{c}
- 16 m^4_{c} (14 \hat{s} + 9 \hat{t}) - 3 \hat{t} (2 \hat{s}^2 }  \\
&&{+ 2 \hat{s} \hat{t} + \hat{t}^2) + 4 m^2_{c} (14 \hat{s}^2 + 20
\hat{s} \hat{t} + 5 \hat{t}^2))]}  ,
\end{eqnarray*}

\par
\begin{eqnarray*}
\frac{d\sigma}{d\hat{t}}\left(gb\to c\overline{c}
\left[{}^3\!P_J^{(8)}\right]b\right)=&& \frac{5 \alpha^3_{s} \pi^2
}{36 \hat{t}^2 (-4 m^3_{c} + m_{c} \hat{t})^3 \hat{s}^2} {[2 m^4_{b}
(28 m^2_{c} - 3 \hat{t}) \hat{t} +
   4 m^2_{b} (224 m^6_{c} - 16 m^4_{c} \hat{t} + 3 \hat{s} \hat{t}^2 }  \\
&&{- 2 m^2_{c} \hat{t} (14 \hat{s} + 5 \hat{t})) + \hat{t} (448 m^6_{c} - 16 m^4_{c} (14 \hat{s} + 9 \hat{t}) - 3 \hat{t} (2 \hat{s}^2 + 2 \hat{s} \hat{t} + \hat{t}^2)} \\
&&{+ 4 m^2_{c} (14 \hat{s}^2 + 20 \hat{s} \hat{t} + 5 \hat{t}^2))]}
,
\end{eqnarray*}

\begin{eqnarray*}
\frac{d\sigma}{d\hat{t}}\left(gb\to b\overline{b}
\left[{}^1\!S_0^{(1)}\right]b\right)=&& \frac{-4 \alpha^3_{s}
\pi^2}{81 m_{b} (m^2_{b} - \hat{s})^4 \hat{t}^2 (-4 m^2_{b} +
\hat{t})^2 (-5 m^2_{b} + \hat{s} + \hat{t})^4 \hat{s}^2}
{ [180000 m^{22}_{b} - 150 m^{20}_{b} }  \\
&&{(5760 \hat{s} - 9527 \hat{t}) + 4 m^{18}_{b} (424800 \hat{s}^2 -
915255 \hat{s} \hat{t} - 908548 \hat{t}^2) +
   \hat{s}^2 \hat{t} (2 \hat{s}^2
} \\
&&{+ 2 \hat{s} \hat{t} + \hat{t}^2) (-3 \hat{s}^3 - 6 \hat{s}^2
\hat{t} + \hat{s} \hat{t}^2 + 4 \hat{t}^3)^2
   + m^{16}_{b} (-1762560 \hat{s}^3 + 3317322 \hat{s}^2 \hat{t}
} \\
&&{+ 4092858 \hat{s} \hat{t}^2 + 3266069 \hat{t}^3) + 4 m^{14}_{b}
(259632 \hat{s}^4 - 394884 \hat{s}^3 \hat{t}
   - 863500 \hat{s}^2 \hat{t}^2
} \\
&&{- 327052 \hat{s} \hat{t}^3 - 374113 \hat{t}^4) - 4 m^{12}_{b}
(88128 \hat{s}^5 - 206589 \hat{s}^4 \hat{t}
   - 287598 \hat{s}^3 \hat{t}^2
} \\
&&{- 325069 \hat{s}^2 \hat{t}^3 + 18200 \hat{s} \hat{t}^4 - 98252
\hat{t}^5) +
   4 m^{10}_{b} (16992 \hat{s}^6 - 122202 \hat{s}^5 \hat{t}
} \\
&&{- 106508 \hat{s}^4 \hat{t}^2 + 3416 \hat{s}^3 \hat{t}^3 -
     95855 \hat{s}^2 \hat{t}^4 + 34362 \hat{s} \hat{t}^5 - 15623 \hat{t}^6) + m^8_{b}
} \\
&&{  (-6912 \hat{s}^7 + 198756 \hat{s}^6 \hat{t} + 253116 \hat{s}^5
\hat{t}^2 - 143898 \hat{s}^4 \hat{t}^3 -
     90600 \hat{s}^3 \hat{t}^4
} \\
&&{+ 132856 \hat{s}^2 \hat{t}^5 - 28370 \hat{s} \hat{t}^6 + 6313
\hat{t}^7) +
   4 m^6_{b} (72 \hat{s}^8 - 12132 \hat{s}^7 \hat{t} - 24340 \hat{s}^6 \hat{t}^2
} \\
&&{+ 2724 \hat{s}^5 \hat{t}^3 + 23713 \hat{s}^4 \hat{t}^4 + 1300
\hat{s}^3 \hat{t}^5 - 9188 \hat{s}^2 \hat{t}^6
   + 177 \hat{s} \hat{t}^7 - 106 \hat{t}^8)
} \\
&&{-4 m^2_{b} \hat{s} \hat{t} (135 \hat{s}^8 + 528 \hat{s}^7 \hat{t}
+ 528 \hat{s}^6 \hat{t}^2 - 271 \hat{s}^5 \hat{t}^3 -
     690 \hat{s}^4 \hat{t}^4 - 215 \hat{s}^3 \hat{t}^5
} \\
&&{+ 187 \hat{s}^2 \hat{t}^6 + 126 \hat{s} \hat{t}^7 + 8 \hat{t}^8)
+
   2 m^4_{b} \hat{t} (3465 \hat{s}^8 + 10044 \hat{s}^7 \hat{t} + 4450 \hat{s}^6 \hat{t}^2
} \\
&&{- 9384 \hat{s}^5 \hat{t}^3 -
     7888 \hat{s}^4 \hat{t}^4 + 1994 \hat{s}^3 \hat{t}^5 + 3007 \hat{s}^2 \hat{t}^6 + 188 \hat{s} \hat{t}^7
   + 8 \hat{t}^8)] } ,
\end{eqnarray*}

\begin{eqnarray*}
\frac{d\sigma}{d\hat{t}}\left(gb\to b\overline{b}
\left[{}^1\!S_0^{(8)}\right]b\right)=&& \frac{ - \alpha^3_{s} \pi^2
}{ 324 m_{b} (m^2_{b} - \hat{s})^4 \hat{t}^2 (-4 m^2_{b}
   + \hat{t})^2 (-5 m^2_{b} + \hat{s} + \hat{t})^4 \hat{s}^2}
{ [ 900000 m^{22}_{b} - 750 m^{20}_{b} } \\
&&{(5760 \hat{s} + 2569 \hat{t}) +
   4 m^{18}_{b} (2124000 \hat{s}^2 + 1514925 \hat{s} \hat{t} - 292642 \hat{t}^2) +
   m^{16}_{b}
} \\
&&{(-8812800 \hat{s}^3 - 6758670 \hat{s}^2 \hat{t} + 3117666 \hat{s}
\hat{t}^2 + 1530353 \hat{t}^3) +
   8 m^{14}_{b}
} \\
&&{(649080 \hat{s}^4 + 282870 \hat{s}^3 \hat{t} - 797714 \hat{s}^2
\hat{t}^2 - 401132 \hat{s} \hat{t}^3 -
     72893 \hat{t}^4) + 4 m^{12}_{b}
} \\
&&{(-440640 \hat{s}^5 + 432465 \hat{s}^4 \hat{t} + 1876374 \hat{s}^3
\hat{t}^2 +
     1308895 \hat{s}^2 \hat{t}^3 + 231988 \hat{s} \hat{t}^4
} \\
&&{+ 29447 \hat{t}^5) +
   4 m^{10}_{b} (84960 \hat{s}^6 - 533250 \hat{s}^5 \hat{t} - 1391312 \hat{s}^4 \hat{t}^2
  - 1158952 \hat{s}^3 \hat{t}^3
} \\
&&{- 439340 \hat{s}^2 \hat{t}^4 - 33147 \hat{s} \hat{t}^5 - 3587
   \hat{t}^6) + \hat{s}^2 \hat{t} (\hat{s} + \hat{t})^2 (90 \hat{s}^6 + 270 \hat{s}^5
   \hat{t}
} \\
&&{+ 363 \hat{s}^4 \hat{t}^2 + 276 \hat{s}^3 \hat{t}^3 + 125
   \hat{s}^2 \hat{t}^4 + 32 \hat{s}
   \hat{t}^5 + 4 \hat{t}^6) + m^8_{b} (-34560 \hat{s}^7
} \\
&&{
   + 976500 \hat{s}^6 \hat{t}
 + 2600172 \hat{s}^5 \hat{t}^2 + 2623254 \hat{s}^4 \hat{t}^3 +
   1433016 \hat{s}^3 \hat{t}^4 + 381832 \hat{s}^2 \hat{t}^5
} \\
&&{
   + 9382 \hat{s} \hat{t}^6
   + 1141 \hat{t}^7) + 4 m^6_{b} (360 \hat{s}^8 - 60660 \hat{s}^7
   \hat{t} - 186316 \hat{s}^6 \hat{t}^2
} \\
&&{- 233112 \hat{s}^5 \hat{t}^3 -165986 \hat{s}^4 \hat{t}^4 - 72110
   \hat{s}^3 \hat{t}^5 - 13166
   \hat{s}^2 \hat{t}^6 - 78 \hat{s} \hat{t}^7 - 19 \hat{t}^8)
} \\
&&{+ 2 m^4_{b} \hat{t} (17325 \hat{s}^8 + 63180 \hat{s}^7 \hat{t} +
   97678 \hat{s}^6 \hat{t}^2 + 87168
   \hat{s}^5 \hat{t}^3 + 50438 \hat{s}^4 \hat{t}^4
} \\
&&{+ 17006 \hat{s}^3 \hat{t}^5 + 1993
   \hat{s}^2 t^6 + 32 \hat{s} \hat{t}^7 + 2 \hat{t}^8) -
   4 m^2_{b} \hat{s} \hat{t} (675 \hat{s}^8 + 2910 \hat{s}^7 \hat{t}
} \\
&&{+ 5448 \hat{s}^6 \hat{t}^2
   + 5912 \hat{s}^5 \hat{t}^3
   + 4167 \hat{s}^4 \hat{t}^4 + 1915 \hat{s}^3 \hat{t}^5 + 484
   \hat{s}^2 \hat{t}^6 + 39 \hat{s} \hat{t}^7 + 2 \hat{t}^8)] } ,
\end{eqnarray*}

\begin{eqnarray*}
\frac{d\sigma}{d\hat{t}}\left(gb\to b\overline{b}
\left[{}^3\!S_1^{(1)}\right]b\right)=&& \frac{-32 \alpha^3_{s}
\pi^2}{243 m_{b} (m^2_{b} - \hat{s})^4 (m^2_{b} - \hat{u})^4 (-2
m^2_{b}
   + \hat{s} + \hat{u})^2 \hat{s}^2} {[5466 m^{18}_{b} - 18668 m^{16}_{b} (\hat{s}
} \\
&&{+ \hat{u}) - 2 \hat{s}^2 \hat{u}^2 (\hat{s} + \hat{u})^3 (9
   \hat{s}^2 + 16 \hat{s} \hat{u} + 9 \hat{u}^2) +
   m^{14}_{b} (23389 \hat{s}^2 + 62070 \hat{s} \hat{u}
} \\
&&{+ 23389 \hat{u}^2) - m^{12}_{b} (14251 \hat{s}^3 + 73585
   \hat{s}^2 \hat{u} + 73585 \hat{s} \hat{u}^2 + 14251 \hat{u}^3)
} \\
&&{+ m^{10}_{b} (4609 \hat{s}^4 + 43259 \hat{s}^3 \hat{u} + 74588
   \hat{s}^2 \hat{u}^2 + 43259 \hat{s} \hat{u}^3 + 4609 \hat{u}^4)
} \\
&&{- m^8_{b} (807 \hat{s}^5 + 14717 \hat{s}^4 \hat{u} + 35184
   \hat{s}^3 \hat{u}^2 + 35184 \hat{s}^2 \hat{u}^3 +
     14717 \hat{s} \hat{u}^4
} \\
&&{+ 807 \hat{u}^5)
   + m^2_{b} \hat{s} \hat{u} (20 \hat{s}^6 + 228 \hat{s}^5 \hat{u} +
   701 \hat{s}^4 \hat{u}^2 + 984 \hat{s}^3 \hat{u}^3
} \\
&&{+ 701 \hat{s}^2 \hat{u}^4 + 228 \hat{s}
   \hat{u}^5 + 20 \hat{u}^6)
   + m^6_{b} (72 \hat{s}^6 + 3179 \hat{s}^5 \hat{u} + 9455 \hat{s}^4
   \hat{u}^2
} \\
&&{+ 12444 \hat{s}^3 \hat{u}^3 + 9455 \hat{s}^2 \hat{u}^4 +
     3179 \hat{s} \hat{u}^5
   + 72 \hat{u}^6) - m^4_{b} (2 \hat{s}^7 + 406 \hat{s}^6 \hat{u}
} \\
&&{+1753 \hat{s}^5 \hat{u}^2
   + 3043 \hat{s}^4 \hat{u}^3 + 3043 \hat{s}^3 \hat{u}^4
   + 1753 \hat{s}^2 \hat{u}^5 + 406 \hat{s} \hat{u}^6 + 2
   \hat{u}^7)]} ,
\end{eqnarray*}

\begin{eqnarray*}
\frac{d\sigma}{d\hat{t}}\left(gb\to b\overline{b}
\left[{}^3\!S_1^{(8)}\right]b\right)=&& \frac{- \alpha^3_{s}
\pi^2}{1944 m^3_{b} (m^2_{b} - s)^4 (m^2_{b} - \hat{u})^4 (-2
m^2_{b} + \hat{s} + \hat{u})^2 \hat{s}^2} {[66174 m^{20}_{b} -
226934 m^{18}_{b}
}  \\
&&{(\hat{s} + \hat{u}) + m^{16}_{b} (260287 \hat{s}^2 + 806952
    \hat{s} \hat{u} + 260287 \hat{u}^2) -
    10 m^{14}_{b} (10975 \hat{s}^3
} \\
&&{+ 96181 \hat{s}^2 \hat{u} + 96181 \hat{s} \hat{u}^2
   + 10975 \hat{u}^3) +
   9 \hat{s}^3 \hat{u}^3 (4 \hat{s}^4 - \hat{s}^3 \hat{u} + 8 \hat{s}^2 \hat{u}^2 - \hat{s} \hat{u}^3
} \\
&&{+ 4 \hat{u}^4) + 2 m^{12}_{b} (6155 \hat{s}^4 + 216310 \hat{s}^3
   \hat{u} + 584812 \hat{s}^2 \hat{u}^2 + 216310 \hat{s} \hat{u}^3
} \\
&&{ +6155 \hat{u}^4) - 2 m^2_{b} \hat{s}^2 \hat{u}^2 (90 \hat{s}^5 +
   493 \hat{s}^4 \hat{u} + 468 \hat{s}^3 \hat{u}^2 +
   468 \hat{s}^2 \hat{u}^3 + 493 \hat{s} \hat{u}^4
} \\
&&{+ 90 \hat{u}^5) -
   2 m^{10}_{b} (1548 \hat{s}^5 + 27601 \hat{s}^4 \hat{u} + 266181 \hat{s}^3 \hat{u}^2 + 266181 \hat{s}^2 \hat{u}^3
} \\
&&{+27601 \hat{s} \hat{u}^4 + 1548 \hat{u}^5) + 2 m^4_{b} \hat{s}
   \hat{u} (94 \hat{s}^6 + 534 \hat{s}^5 \hat{u} +
   3748 \hat{s}^4 \hat{u}^2 - 477 \hat{s}^3 \hat{u}^3
} \\
&&{+ 3748 \hat{s}^2 \hat{u}^4 + 534 \hat{s} \hat{u}^5 + 94
   \hat{u}^6) + m^8_{b} (285 \hat{s}^6 + 9920 \hat{s}^5 \hat{u} + 60701 \hat{s}^4 \hat{u}^2
} \\
&&{+ 245736 \hat{s}^3 \hat{u}^3 + 60701 \hat{s}^2 \hat{u}^4 + 9920
   \hat{s} \hat{u}^5 + 285 \hat{u}^6) -
   2 m^6_{b} (22 \hat{s}^7 + 563 \hat{s}^6 \hat{u}
} \\
&&{+ 4424 \hat{s}^5 \hat{u}^2 + 13715 \hat{s}^4 \hat{u}^3 + 13715
   \hat{s}^3 \hat{u}^4 + 4424 \hat{s}^2 \hat{u}^5 + 563 \hat{s}
   \hat{u}^6 + 22 \hat{u}^7) ]} ,
\end{eqnarray*}

\begin{eqnarray*}
\frac{d\sigma}{d\hat{t}}\left(gb\to b\overline{b}
\left[{}^1\!P_1^{(1)}\right]b\right)=&& \frac{64 \alpha^3_{s}
\pi^2}{243 (m^2_{b} - \hat{s})^5 (-5 m^2_{b} + \hat{s} + \hat{t})^5
(4 m^3_{b} - m_{b} \hat{t})^3 \hat{s}^2} { [2227744 m^{24}_{b} - 2
m^{22}_{b}
} \\
&&{
  (10071936 \hat{s} + 1099435 \hat{t}) -
   \hat{s}^3 t^4 (\hat{s} + \hat{t})^3 (2 \hat{s}^2 + 2 \hat{s} \hat{t} + 9 \hat{t}^2) +
   m^{20}_{b}
} \\
&&{(-6754944 \hat{s}^2 + 28431108 \hat{s} \hat{t} + 367984
   \hat{t}^2) + m^{18}_{b} (12743424 \hat{s}^3
} \\
&&{+ 25261690 \hat{s}^2 \hat{t} - 16277654 \hat{s} \hat{t}^2 +
   488169 \hat{t}^3) - 2 m^{16}_{b} (3126432 \hat{s}^4
} \\
&&{+ 13119432 \hat{s}^3 \hat{t} + 14986264 \hat{s}^2 \hat{t}^2 -
   2170280 \hat{s} \hat{t}^3 + 183541 \hat{t}^4) + 4 m^{14}_{b}
} \\
&&{(409536 \hat{s}^5 + 2567733 \hat{s}^4 \hat{t} + 5640894 \hat{s}^3
   \hat{t}^2 + 4564301 \hat{s}^2 \hat{t}^3 - 49258 \hat{s} \hat{t}^4
} \\
&&{+ 32054 \hat{t}^5) + m^2_{b} \hat{s}^2 \hat{t} (\hat{s} +
   \hat{t})^2 (2 \hat{s}^6 + 6 \hat{s}^5 \hat{t} + 35 \hat{s}^4
   \hat{t}^2 + 108 \hat{s}^3 \hat{t}^3 + 249 \hat{s}^2 \hat{t}^4
} \\
&&{+ 334 \hat{s} \hat{t}^5 + 27 \hat{t}^6) - m^{12}_{b} (257664
   \hat{s}^6 + 2250792 \hat{s}^5 \hat{t} + 7210544 \hat{s}^4 \hat{t}^2
} \\
&&{+ 10902208 \hat{s}^3 \hat{t}^3 + 6680880 \hat{s}^2 \hat{t}^4 +
   222148 \hat{s} \hat{t}^5 + 27053 \hat{t}^6) + 2 m^{10}_{b}
} \\
&&{(11904 \hat{s}^7 + 152098 \hat{s}^6 \hat{t} + 651630 \hat{s}^5
   \hat{t}^2 + 1443207 \hat{s}^4 \hat{t}^3 + 1656100 \hat{s}^3 \hat{t}^4
} \\
&&{+ 780410 \hat{s}^2 \hat{t}^5 + 36206 \hat{s} \hat{t}^6 + 1776
   \hat{t}^7) + m^6_{b} \hat{t} (1474 \hat{s}^8 + 10168 \hat{s}^7 \hat{t}
} \\
&&{+ 36644 \hat{s}^6 \hat{t}^2 + 87208 \hat{s}^5 \hat{t}^3
   +125184 \hat{s}^4 \hat{t}^4 + 88392 \hat{s}^3 \hat{t}^5 + 22168
   \hat{s}^2 \hat{t}^6
} \\
&&{+ 836 \hat{s} \hat{t}^7 + 9 \hat{t}^8) - m^4_{b} \hat{s} \hat{t}
   (60 \hat{s}^8 + 448 \hat{s}^7 \hat{t} + 1888 \hat{s}^6 \hat{t}^2 +
   5568 \hat{s}^5 \hat{t}^3
} \\
&&{+ 11620 \hat{s}^4 \hat{t}^4 + 14011 \hat{s}^3 \hat{t}^5 + 7614
   \hat{s}^2 \hat{t}^6 + 1182 \hat{s} \hat{t}^7 + 27 \hat{t}^8) -
   m^8_{b} (992 \hat{s}^8
} \\
&&{+ 26384 \hat{s}^7 \hat{t} + 143952 \hat{s}^6 \hat{t}^2 + 424880
   \hat{s}^5 \hat{t}^3 + 736548 \hat{s}^4 \hat{t}^4
   + 663440 \hat{s}^3 \hat{t}^5
} \\
&&{+ 235347 \hat{s}^2 \hat{t}^6 + 10823 \hat{s} \hat{t}^7 + 269
   \hat{t}^8) ]
} ,
\end{eqnarray*}

\begin{eqnarray*}
\frac{d\sigma}{d\hat{t}}\left(gb\to b\overline{b}
\left[{}^1\!P_1^{(8)}\right]b\right) = && \frac{\alpha^3_{s}
\pi^2}{972 (m^2_{b} - \hat{s})^5 \hat{t}^2 (-5 m^2_{b} + \hat{s} +
\hat{t})^5 (4 m^3_{b} - m_{b} \hat{t})^3 \hat{s}^2} { [ 32400000
m^{28}_{b} -
5400 m^{26}_{b} }  \\
&&{} (36000 \hat{s} +
  57209 \hat{t}) + 2 m^{24}_{b} (249480000 \hat{s}^2 + 561226320
  \hat{s} \hat{t} + 23814083 \hat{t}^2)
\\
&&{}  - 2 m^{22}_{b} (357696000 \hat{s}^3 + 843783912 \hat{s}^2
  \hat{t} + 267419628 \hat{s} \hat{t}^2 - 77026763 \hat{t}^3)
\\
&&{}  + m^{20}_{b} (628819200 \hat{s}^4 + 1367173728 \hat{s}^3
\hat{t} +
     912846660 \hat{s}^2 \hat{t}^2 - 48175560 \hat{s} \hat{t}^3
\\
&&{}
     - 99056939 \hat{t}^4) +
   m^{18}_{b} (-351226368 \hat{s}^5 - 604100808 \hat{s}^4 \hat{t} - 504111528 \hat{s}^3 \hat{t}^2
\\
&&{}
   +
     60998614 \hat{s}^2 \hat{t}^3 + 89194312 \hat{s} \hat{t}^4 + 28705941 \hat{t}^5) +
   m^{16}_{b} (125763840 \hat{s}^6
\\
&&{}
   + 69346368 \hat{s}^5 \hat{t} - 243821166 \hat{s}^4 \hat{t}^2 -
     449357484 \hat{s}^3 \hat{t}^3 - 283946945 \hat{s}^2 \hat{t}^4
\\
&&{}
     - 37003547 \hat{s} \hat{t}^5 - 5575172 \hat{t}^6) -
   \hat{s}^3 \hat{t}^2 (\hat{s} + \hat{t})^3 (162 \hat{s}^6 + 486 \hat{s}^5 \hat{t} + 693 \hat{s}^4 \hat{t}^2
\\
&&{}
   + 576 \hat{s}^3 \hat{t}^3 +
     323 \hat{s}^2 \hat{t}^4 + 116 \hat{s} \hat{t}^5 + 36 \hat{t}^6) +
   4 m^{14}_{b} (-7153920 \hat{s}^7
   + 20626056 \hat{s}^6 \hat{t}
\\
&&{}
   + 112036932 \hat{s}^5 \hat{t}^2 + 207447531 \hat{s}^4 \hat{t}^3 + 131086836 \hat{s}^3 \hat{t}^4
     + 43878881 \hat{s}^2 \hat{t}^5
\\
&&{}
     +3359759 \hat{s} \hat{t}^6 + 243128 \hat{t}^7) + m^{12}_{b} (3991680
     \hat{s}^8 - 55432512 \hat{s}^7 \hat{t}
\\
&&{}
     - 267148776 \hat{s}^6 \hat{t}^2 - 604619712 \hat{s}^5 \hat{t}^3 -
587249390 \hat{s}^4 \hat{t}^4 -
     245568940 \hat{s}^3 \hat{t}^5
\\
&&{}
     - 61736796 \hat{s}^2 \hat{t}^6 - 3743122 \hat{s} \hat{t}^7 - 151151 \hat{t}^8) +
   m^2_{b} \hat{s}^2 \hat{t} (\hat{s} + \hat{t})^2 (648 \hat{s}^8
\\
&&{}
   + 8424 \hat{s}^7 \hat{t} + 29366 \hat{s}^6 \hat{t}^2 +
     50244 \hat{s}^5 \hat{t}^3 + 47759 \hat{s}^4 \hat{t}^4 + 26802 \hat{s}^3 \hat{t}^5 + 8889 \hat{s}^2 \hat{t}^6
\\
&&{}
     +
     1840 \hat{s} \hat{t}^7 + 108 \hat{t}^8) + m^{10}_{b} (-311040 \hat{s}^9 + 17461656 \hat{s}^8 \hat{t} +
     91385904 \hat{s}^7 \hat{t}^2
\\
&&{}
     + 243458012 \hat{s}^6 \hat{t}^3 + 322380960 \hat{s}^5 \hat{t}^4 +
     207346758 \hat{s}^4 \hat{t}^5 + 66243932 \hat{s}^3 \hat{t}^6
\\
&&{}
     + 13294060 \hat{s}^2 \hat{t}^7 + 642412 \hat{s} \hat{t}^8 +
     16665 \hat{t}^9) + m^8_{b} (10368 \hat{s}^{10} - 3252960 \hat{s}^9 \hat{t}
\\
&&{}
     - 19662902 \hat{s}^8 \hat{t}^2 -
     60570920 \hat{s}^7 \hat{t}^3 - 99701250 \hat{s}^6 \hat{t}^4 - 88396106 \hat{s}^5 \hat{t}^5
\\
&&{}
      - 42116100 \hat{s}^4 \hat{t}^6 - 10893998 \hat{s}^3 \hat{t}^7 -
     1728483 \hat{s}^2 \hat{t}^8 - 64271 \hat{s} \hat{t}^9 -
     1112 \hat{t}^{10})
\\
&&{}
     + 2 m^6_{b} \hat{t} (183384 \hat{s}^{10} + 1338660 \hat{s}^9 \hat{t} + 4750079 \hat{s}^8 \hat{t}^2 +
     9250520 \hat{s}^7 \hat{t}^3
\\
&&{}
     + 10317034 \hat{s}^6 \hat{t}^4 + 6712106 \hat{s}^5 \hat{t}^5 + 2525028 \hat{s}^4 \hat{t}^6 +
     540096 \hat{s}^3 \hat{t}^7 + 65441 \hat{s}^2 \hat{t}^8
\\
&&{}
     + 1852 \hat{s} \hat{t}^9 + 18 \hat{t}^{10}) -
   m^4_{b} \hat{s} \hat{t} (23328 \hat{s}^{10} + 220572 \hat{s}^9 \hat{t} + 904920 \hat{s}^8 \hat{t}^2
\\
&&{}
    + 2024407 \hat{s}^7 \hat{t}^3 +
     2686780 \hat{s}^6 \hat{t}^4 + 2197140 \hat{s}^5 \hat{t}^5 + 1116238 \hat{s}^4 \hat{t}^6 + 346213 \hat{s}^3 \hat{t}^7
\\
&&{}
     +62406 \hat{s}^2 \hat{t}^8 + 5448 \hat{s} \hat{t}^9 + 108 \hat{t}^{10}) ] ,
\end{eqnarray*}

\begin{eqnarray*}
\frac{d\sigma}{d\hat{t}}\left(gb\to b\overline{b}
\left[{}^3\!P_J^{(1)}\right]b\right)= && \frac{4 \alpha^3_{s}
\pi^2}{81 (m^2_{b} - \hat{s})^5 \hat{t}^2 (-5 m^2_{b} + \hat{s} +
\hat{t})^5 (4 m^3_{b} - m_{b} \hat{t})^3 \hat{s}^2}
{[ 25200000 m^{28}_{b} - 600 m^{26}_{b}} \\
&&{ (252000 \hat{s} + 346111 \hat{t}) -
   3 \hat{s}^3 \hat{t}^2 (\hat{s} + \hat{t})^3 (3 \hat{s}^2 + 3 \hat{s} \hat{t} - 4 \hat{t}^2)^2 (2 \hat{s}^2 + 2 \hat{s} \hat{t} + \hat{t}^2)
}\\
&&{}
   + 18 m^{24}_{b} (21560000 \hat{s}^2 + 26837680 \hat{s} \hat{t} +
   6879209 \hat{t}^2) - 6 m^{22}_{b} (92736000 \hat{s}^3
\\
&&{}
   + 58438984 \hat{s}^2 \hat{t} + 132101836 \hat{s} \hat{t}^2 - 31650279 \hat{t}^3) +
   m^{20}_{b} (489081600 \hat{s}^4
\\
&&{}
   + 52223904 \hat{s}^3 \hat{t} - 596240052 \hat{s}^2 \hat{t}^2 +
   475090104 \hat{s} \hat{t}^3 - 316136919 \hat{t}^4)
\\
&&{}
   + 3 m^{18}_{b} (-91058688 \hat{s}^5 + 34245912 \hat{s}^4
   \hat{t} + 252722648 \hat{s}^3 \hat{t}^2 + 543856590 \hat{s}^2 \hat{t}^3
\\
&&{}
   + 29438668 \hat{s} \hat{t}^4 + 71917105 \hat{t}^5) +
   m^{16}_{b} (97816320 \hat{s}^6 - 160017984 \hat{s}^5 \hat{t}
\\
&&{}
   - 491228090 \hat{s}^4 \hat{t}^2 -
   877626916 \hat{s}^3 \hat{t}^3 - 1687844825 \hat{s}^2 \hat{t}^4 - 268763655 \hat{s} \hat{t}^5
\\
&&{}
   - 87840460 \hat{t}^6) + 4 m^{14}_{b} (-5564160 \hat{s}^7 +
   30448824 \hat{s}^6 \hat{t} + 75718764 \hat{s}^5 \hat{t}^2
\\
&&{}
   + 66864981 \hat{s}^4 \hat{t}^3 + 140914224 \hat{s}^3 \hat{t}^4
   + 246581099 \hat{s}^2 \hat{t}^5 + 40404191 \hat{s} \hat{t}^6
\\
&&{}
   + 5840348 \hat{t}^7) +
   m^{12}_{b} (3104640 \hat{s}^8 - 54127296 \hat{s}^7 \hat{t} - 151809336 \hat{s}^6 \hat{t}^2
\\
&&{}
   - 114986592 \hat{s}^5 \hat{t}^3 - 65176694 \hat{s}^4 \hat{t}^4
   - 238851724 \hat{s}^3 \hat{t}^5 - 359985180 \hat{s}^2 \hat{t}^6
\\
&&{}
   - 52213906 \hat{s} \hat{t}^7 - 4131047 \hat{t}^8) +
   m^2_{b} \hat{s}^2 \hat{t} (\hat{s} + \hat{t})^2 (504 \hat{s}^8 + 3960 \hat{s}^7 \hat{t}
\\
&&{}
   + 8522 \hat{s}^6 \hat{t}^2 + 3336 \hat{s}^5 \hat{t}^3 -
   4911 \hat{s}^4 \hat{t}^4 - 1942 \hat{s}^3 \hat{t}^5 + 791 \hat{s}^2 \hat{t}^6 + 416 \hat{s} \hat{t}^7
\\
&&{}
   + 1552 \hat{t}^8) + m^{10}_{b} (-241920 \hat{s}^9 + 14973288
   \hat{s}^8 \hat{t} +
   51568272 \hat{s}^7 \hat{t}^2
\\
&&{}
   + 52527524 \hat{s}^6 \hat{t}^3 + 13171944 \hat{s}^5 \hat{t}^4 +
   8800346 \hat{s}^4 \hat{t}^5 + 67284444 \hat{s}^3 \hat{t}^6
\\
&&{}
   + 85100380 \hat{s}^2 \hat{t}^7 + 10250336 \hat{s} \hat{t}^8 +
   471419 \hat{t}^9) + m^8_{b} (8064 \hat{s}^{10}
\\
&&{}
   - 2633760 \hat{s}^9 \hat{t}
   - 11282354 \hat{s}^8 \hat{t}^2 - 15962360 \hat{s}^7 \hat{t}^3 -
   7053666 \hat{s}^6 \hat{t}^4
\\
&&{}
   + 1332094 \hat{s}^5 \hat{t}^5 - 874708 \hat{s}^4 \hat{t}^6
   - 12363406 \hat{s}^3 \hat{t}^7 -
   13033959 \hat{s}^2 \hat{t}^8
\\
&&{}
   - 1228903 \hat{s} \hat{t}^9 - 31560 \hat{t}^{10}) +
   2 m^6_{b} \hat{t} (144360 \hat{s}^{10} + 773580 \hat{s}^9 \hat{t}
\\
&&{}
   + 1467393 \hat{s}^8 \hat{t}^2 + 1038192 \hat{s}^7 \hat{t}^3 +
   56862 \hat{s}^6 \hat{t}^4 - 231222 \hat{s}^5 \hat{t}^5 + 62812 \hat{s}^4 \hat{t}^6
\\
&&{}
   + 716948 \hat{s}^3 \hat{t}^7 + 623079 \hat{s}^2 \hat{t}^8 + 41600 \hat{s} \hat{t}^9 + 472
   \hat{t}^{10}) - m^4_{b} \hat{s} \hat{t} (18144 \hat{s}^{10}
\\
&&{}
   + 124628 \hat{s}^9 \hat{t} + 311800 \hat{s}^8 \hat{t}^2 +
   318915 \hat{s}^7 \hat{t}^3 + 73212 \hat{s}^6 \hat{t}^4 - 68716 \hat{s}^5 \hat{t}^5
\\
&&{}
   - 37362 \hat{s}^4 \hat{t}^6 + 24645 \hat{s}^3 \hat{t}^7 +
   98510 \hat{s}^2 \hat{t}^8 + 67280 \hat{s} \hat{t}^9 + 2448 \hat{t}^{10}) ] ,
\end{eqnarray*}

\begin{eqnarray*}
\frac{d\sigma}{d\hat{t}}\left(gb\to b\overline{b}
\left[{}^3\!P_J^{(8)}\right]b\right)=&& \frac{ \alpha^3_{s}
\pi^2}{324 (m^2_{b} - \hat{s})^5 \hat{t}^2 (-5 m^2_{b} + \hat{s} +
\hat{t})^5 (4 m^3_{b} - m_{b} \hat{t})^3 \hat{s}^2} { [126000000
m^{28}_{b} }
\\
&&{}
   - 21000 m^{26}_{b} (36000 \hat{s} - 5111 \hat{t}) +
   6 m^{24}_{b} (323400000 \hat{s}^2 - 10772400 \hat{s} \hat{t}
\\
&&{}
   - 179385097 \hat{t}^2) -
   6 m^{22}_{b} (463680000 \hat{s}^3 + 66541160 \hat{s}^2 \hat{t} - 356589988 \hat{s} \hat{t}^2
\\
&&{}
   - 199413487 \hat{t}^3) + 3 m^{20}_{b} (815136000 \hat{s}^4 + 161689440
   \hat{s}^3 \hat{t} - 713740812 \hat{s}^2 \hat{t}^2
\\
&&{}
   - 765731304 \hat{s} \hat{t}^3 - 210248369 \hat{t}^4) -
   3 m^{18}_{b} (455293440 \hat{s}^5 - 60499320 \hat{s}^4 \hat{t}
\\
&&{}
   - 839359544 \hat{s}^3 \hat{t}^2 -
   998696926 \hat{s}^2 \hat{t}^3 - 412742996 \hat{s} \hat{t}^4 - 69091505 \hat{t}^5) +
   m^{16}_{b}
\\
&&{}
   ( 489081600 \hat{s}^6 - 695580480 \hat{s}^5 \hat{t} -
   3191554274 \hat{s}^4 \hat{t}^2 -
   3864588148 \hat{s}^3 \hat{t}^3
\\
&&{}
   - 2108766941 \hat{s}^2 \hat{t}^4 - 451377411 \hat{s} \hat{t}^5
   - 49514524 \hat{t}^6 ) - 3 \hat{s}^3 \hat{t}^2 ( \hat{s} + \hat{t})^3
   ( 90 \hat{s}^6
\\
&&{}
   + 270 \hat{s}^5 \hat{t} + 363 \hat{s}^4 \hat{t}^2 +
   276 \hat{s}^3 \hat{t}^3 + 125 \hat{s}^2 \hat{t}^4 + 32 \hat{s} \hat{t}^5 + 4 \hat{t}^6 ) +
   4 m^{14}_{b}
\\
&&{}
   ( -27820800 \hat{s}^7 + 147889560 \hat{s}^6 \hat{t} + 590044860
   \hat{s}^5 \hat{t}^2 +
   991176813 \hat{s}^4 \hat{t}^3
\\
&&{}
   + 673013160 \hat{s}^3 \hat{t}^4 + 228128459 \hat{s}^2 \hat{t}^5
   + 31137503 \hat{s} \hat{t}^6 + 2314418 \hat{t}^7 )
\\
&&{}
   + m^{12}_{b} ( 15523200 \hat{s}^8 - 268977600 \hat{s}^7 \hat{t}
   - 1077142104 \hat{s}^6 \hat{t}^2
   - 2165379264 \hat{s}^5 \hat{t}^3
\\
&&{}
   - 2341866062 \hat{s}^4 \hat{t}^4 -
   1031915260 \hat{s}^3 \hat{t}^5 - 254801556 \hat{s}^2 \hat{t}^6 - 25728466 \hat{s} \hat{t}^7
\\
&&{}
   - 1320599 \hat{t}^8 ) +
   m^2_{b} \hat{s}^2 \hat{t} ( \hat{s} + \hat{t} )^2 (2520 \hat{s}^8 + 19800 \hat{s}^7 \hat{t} + 51098 \hat{s}^6 \hat{t}^2
\\
&&{}
   + 68064 \hat{s}^5 \hat{t}^3 + 64785 \hat{s}^4 \hat{t}^4 + 48194
   \hat{s}^3 \hat{t}^5 + 17783 \hat{s}^2 \hat{t}^6 +
   344 \hat{s} \hat{t}^7 + 388 \hat{t}^8)
\\
&&{}
   + m^{10}_{b} (-1209600 \hat{s}^9 + 74797320 \hat{s}^8 \hat{t} +
   324287952 \hat{s}^7 \hat{t}^2 + 702527876 \hat{s}^6 \hat{t}^3
\\
&&{}
   + 1006700472 \hat{s}^5 \hat{t}^4 +
   755209754 \hat{s}^4 \hat{t}^5 + 232438668 \hat{s}^3 \hat{t}^6 + 46247836 \hat{s}^2 \hat{t}^7
\\
&&{}
   + 3738896 \hat{s} \hat{t}^8 + 131591 \hat{t}^9) + m^8_{b} ( 40320
   \hat{s}^{10} - 13168800 \hat{s}^9 \hat{t} -
   64937498 \hat{s}^8 \hat{t}^2
\\
&&{}
   - 149484440 \hat{s}^7 \hat{t}^3 - 240269802 \hat{s}^6 \hat{t}^4
   - 255755594 \hat{s}^5 \hat{t}^5 - 139606876 \hat{s}^4 \hat{t}^6
\\
&&{}
   - 30968470 \hat{s}^3 \hat{t}^7 -
   5385543 \hat{s}^2 \hat{t}^8 - 359983 \hat{s} \hat{t}^9 - 8100 \hat{t}^{10}) +
   2 m^6_{b} \hat{t}
\\
&&{}
   ( 721800 \hat{s}^{10} + 4187100 \hat{s}^9 \hat{t} + 10660281 \hat{s}^8 \hat{t}^2 +
   17776032 \hat{s}^7 \hat{t}^3 + 22160286 \hat{s}^6 \hat{t}^4
\\
&&{}
   + 17930202 \hat{s}^5 \hat{t}^5 +
   7326376 \hat{s}^4 \hat{t}^6 + 1157204 \hat{s}^3 \hat{t}^7 + 196287 \hat{s}^2 \hat{t}^8
   + 10640 \hat{s} \hat{t}^9
\\
&&{}
   + 118 \hat{t}^{10}) - m^4_{b} \hat{s} \hat{t} (90720 \hat{s}^{10} +
   644420 \hat{s}^9 \hat{t} + 1920040 \hat{s}^8 \hat{t}^2 +
   3393519 \hat{s}^7 \hat{t}^3
\\
&&{}
   + 4354860 \hat{s}^6 \hat{t}^4 + 4215260 \hat{s}^5 \hat{t}^5 +
   2625438 \hat{s}^4 \hat{t}^6 +
   805269 \hat{s}^3 \hat{t}^7 + 84038 \hat{s}^2 \hat{t}^8
\\
&&{}
   + 17240 \hat{s} \hat{t}^9 + 612 \hat{t}^{10} ) ] .
\end{eqnarray*}

\end{appendix}

\end{document}